\documentclass[a4paper,fleqn,usenatbib]{mnras}

\usepackage{mathptmx}

\usepackage[T1]{fontenc}
\usepackage{ae,aecompl}

\usepackage{textcmds}
\usepackage{graphicx}	
\usepackage{float}
\usepackage{dblfloatfix}

\usepackage{amsmath}	
\usepackage{amssymb}	
\usepackage{bm,bbm, mathtools}

\usepackage{xcolor}
\definecolor{ao}{rgb}{0.0, 0.5, 0.0}

\usepackage{array,arydshln} 
\usepackage{booktabs} 
\usepackage{longtable}

\newcolumntype{$}{>{\global\let\currentrowstyle\relax}}
\newcolumntype{^}{>{\currentrowstyle}}

\newcolumntype{L}[1]{>{\raggedright\arraybackslash}p{#1}} 
\newcolumntype{C}[1]{>{\centering\arraybackslash}m{#1}} 
\newcolumntype{R}[1]{>{\raggedleft\arraybackslash}p{#1}} 

\renewcommand{\exp}[1]{\mathrm{exp}\left(#1\right)}
\newcommand\dd{\,\mathrm{d}}

\renewcommand{\eqref}[1]{Eq.~(\ref{#1})}
\newcommand{\tabref}[1]{Tab.~\ref{#1}}
\newcommand{\figref}[1]{Fig.~\ref{#1}}

\title[Motion of chondrules by temperature fluctuations]{The motion of chondrules and other particles in a protoplanetary disc with temperature fluctuations}

\author[Loesche et al.]{
C. Loesche,$^{1}$\thanks{E-mail: christoph.loesche@uni-due.de}
G. Wurm,$^{1}$
T. Kelling,$^{1}$
J. Teiser,$^{1}$
and D.~S. Ebel$^{2}$
\\
$^{1}$Fakult{\"a}t f{\"u}r Physik, Universit{\"a}t Duisburg-Essen, Lotharstr. 1, 47048 Duisburg, Germany\\
$^{2}$Department of Earth and Planetary Science, American Museum of Natural History, New York, NY 10024, USA
}

\date{Accepted XXX. Received YYY; in original form ZZZ}

\pubyear{2016}

\begin{document}
\label{firstpage}
\pagerange{\pageref{firstpage}--\pageref{lastpage}}
\maketitle

\begin{abstract}
We consider the mechanism of photophoretic transport in protoplanetary disks that are optically thick to radiation.
Here, photophoresis is not caused by the central star but by temperature fluctuations that subject suspended solid particles, including chondrules, to non-isotropic thermal radiation within the disk.
These short-lived temperature fluctuations can explain time-of-flight size sorting and general number density enhancements.
The same mechanism will also lead to velocity fluctuations of dust aggregates beyond $100\,\mathrm{m\,s^{-1}}$ for mm-sized particles in protoplanetary disks.
Applying this in future research will change our understanding of the early phases of collisional dust evolution and aggregate growth as particles cross the bouncing barrier and as mass transfer rates are altered.
\end{abstract}




\section{Introduction}
\begin{figure}
\includegraphics[width=0.5\textwidth]{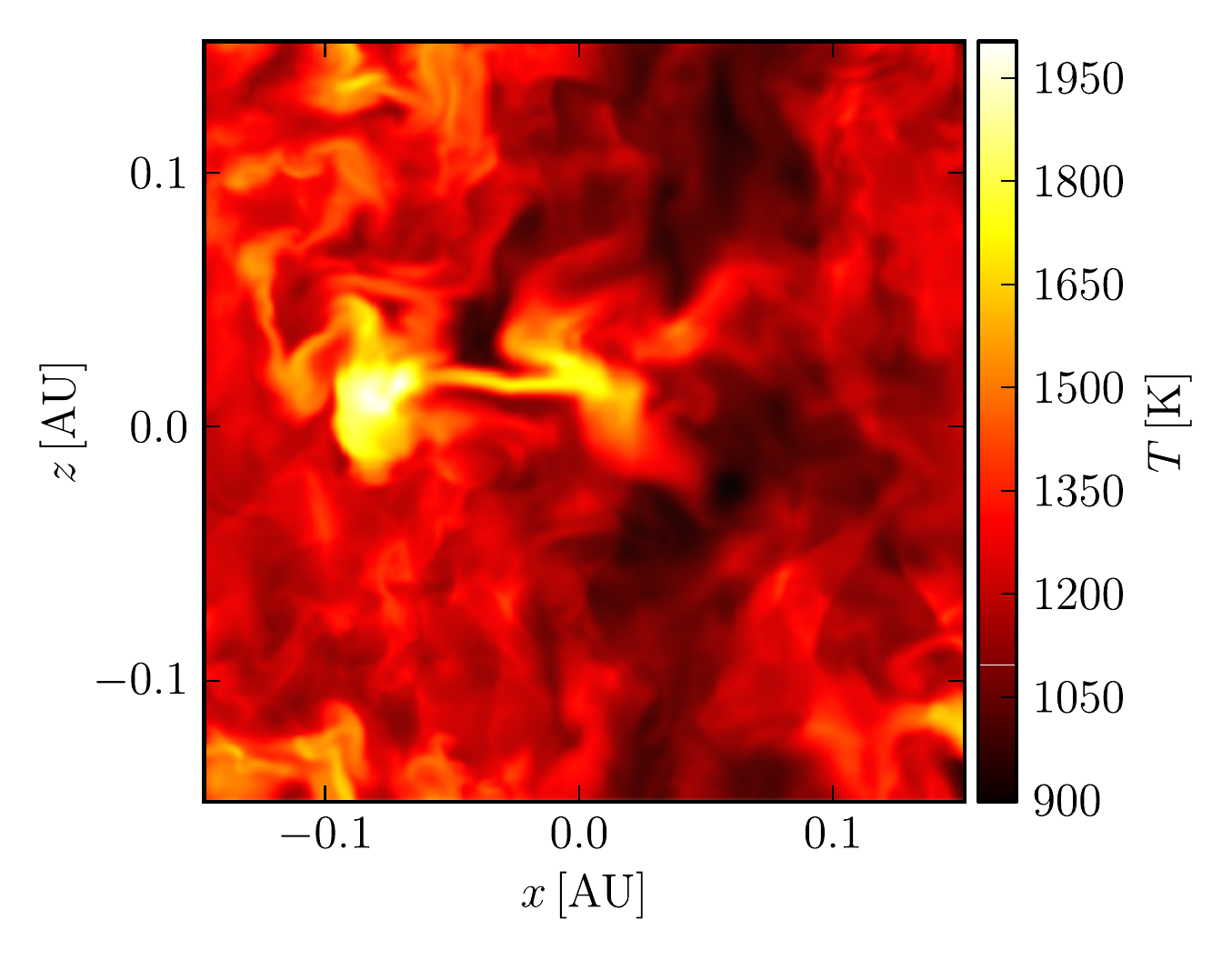}
\caption{Temperature fluctuations in a magnetized 
 disks showing steep temperature gradients. Reprinted from \citet{mcnally2014}.}
\label{fig:tempfluc}
\end{figure}

Chondrules are mm-size particles that participated in the early phases of planet formation.
During their formation {period} of a few million years they experienced strong heating events.
On one hand these events were hot enough to melt the chondrules, reaching temperatures $\ge 1500$ K
\citep{Scott2007}.
On the other hand these heating events must only have occurred intermittently, and regions of temperature hundreds of degrees cooler had to have been close by \citep{Connolly2006}.
This implies that gas and dust temperatures varied on spatially small scales in the solar nebula.
To explain this, local heat sources have to be considered. Heating mechanisms discussed have included shocks \citep{2010ApJ...722.1474M,Hood1991,Ciesla2002,hubbard2015, Morris2012}, jet-flow \citep{Liffman1996}, X-wind \citep{1990LPI....21.1166S,1990LPI....21.1168S,Shu1996}, impacts \citep{Krot2005,Dullemond2014}, or current sheets in the magnetically active region of a disk \citep{Hubbard2012,mcnally2013}.
We do not judge the chondrule forming mechanism in this paper, but consider current sheets illustrative.
They give a picture of a solar nebula with extremely inhomogenous temperatures \citep[][see also \figref{fig:tempfluc}]{mcnally2014}.
Per definition all scenarios result in a local region of the disk that is hot enough to melt chondrules but is itself close to a cooler region of the disk to allow cooling.

Such a temperature distribution in a protoplanetary disk has certain effects.
Usually, protoplanetary disks are considered to be optically thick to wavelengths shorter than infrared.
This indeed holds for radiation traversing the disk between surface and midplane.
Therefore, e.g. stellar radiation cannot reach the midplane.
However, {optical depths for thermal radiation are much lower for the short pathlengths across the boundaries of small-scale temperature fluctuations.
In this case, solids within the disk are subject to anisotropic thermal radiation.
Such directed radiation can push particles from the brighter, warm side to the darker, cool side either by radiation pressure or by interaction with the surrounding gas causing photophoretic motion.
We estimate below that the resulting motion can dominate local particle transport, so this could have far reaching consequences.

\begin{description}
\item[\textbf{Chondrules}] It might explain why chondrules only stay in a hot region for a short time, as they are pushed towards cooler regions.
Several studies have reported that each meteorite can be characterized by a narrow size distribution \citep{Teitler2010, Friedrich2015ChEG...75..419F, Friedrich2015}.
Local transport by photophoresis can sort chondrules according to size.
Moving solids through the solar nebula or other protoplanetary disks will additionally increase the solid number density in some parts of the disk, which could serve as a gravitational seed for later planetesimal or asteroidal formation by gravitational instabilities.
\item[\textbf{Other solids}] Chondrules are only one example of solids of well defined size requiring temperature fluctuations, and the above considerations may also apply to other classes of particles.
Chondrule progenitors were likely assemblies of dust of various size.
Heating events might not always reach high enough peaks to melt such aggregates.
Even so, temperature fluctuations will drive higher particle velocities along the local gradient of thermal radiation.
In fact, as estimated below, this velocity component might  dominate the motion, with magnitudes reaching 100--200 m~s$^{-1}$.
This will change the behavior of solids near bouncing \citep{Zsom2010,Kelling2014}, charging \citep{konopka2005,okuzumi2009}, or fragmentation barriers \citep{guettler2010}, and could affect the predicted course of planetesimal formation.
\end{description}

The detailed evolution of solids is a subject for future work as particle aggregation is a complex story on its own.
In this article we argue that temperature fluctuations are not only a heating mechanism for solid particles but also are of fundamental importance to particle transport, concentration and collisions.
Radiation-driven motion of solids can be significant in a disk even if the disk is optically thick on a global scale.

We begin with a description of typical temperature fluctuations, and then use them to estimate the effects of radiation pressure from temperature fluctuations on particles in a toy model, which are already non-negligible.
We then quantify photophoretic forces on chondrules and dust aggregates, determine drift velocities, and describe sorting processes.
\section{Temperature fluctuations}

In this work we consider a \emph{hot spot} temperature fluctuation. 
It is most directly modeled after the effect of a magnetic energy dissipation event \citep[a current sheet,][]{mcnally2014} but is a reasonably general approximation for the effect of any localized heating event in a protoplanetary disk.
For simplicity we adopt a model where the gas density is constant across the temperature fluctuation.
Our parameter study varies the ratio between optical depth and temperature fluctuation length scale as detailed below.
We model only one-sided temperature transitions in our one-dimensional models.

Gas and dust (particle) temperatures can differ.
Especially in low-pressure regions, coupling between gas and dust might not equilibrate temperatures on suitable timescales. 
While heating mechanisms such as current sheets act on the gas, it is the solid fraction that efficiently radiates thermally. Therefore, the particle temperature is the important one for our purposes. 
However, because we assume that the gas is the heated component, we must determine under what conditions the gas efficiently heats the particles.

To answer this question, we estimate the timescale and final temperature of a suspended dust particle compared to the gas temperature.
This establishes the thermal radiation field, as the gas itself is optically thin.
The radiation field---as modulated by the gas temperature fluctuations---then is responsible for the actual transport and migration of particles like chondrules by photophoresis and radiation pressure.

For this estimate, we solve a heat transfer problem for a spherical dust particle with radius $a$ and temperature $T_{\text{d}}(\mathbf{x},t)$, embedded in a gas of temperature $T_{\text{g}}$, which fluctuates between the initial, cool temperature $T_\text{c}$ and the final, warm temperature $T_\text{w}>T_\text{c}$
\begin{equation}
	\boldsymbol{\nabla}\cdot k\, \boldsymbol{\nabla} T_\text{d} = \rho_\text{d}\,c_p\,\partial_t T_\text{d} \; .
\end{equation}
$c_p$ and $\rho_\text{d}$ are the isobaric heat capacity and the mass density of the dust particle, respectively.
For this spherical problem the spatial domain is $0< |\mathbf{x}| \le a$.
 The boundary condition at the surface is
\begin{equation}
   h\,\left(T_\text{d}-T_\text{g}\right) +\left.  k\frac{\partial T_\text{d}}{\partial \mathbf{n}} \right|_{|\mathbf{x}| = a} = 0 \;, \label{eq:xfer}
\end{equation}
where $k$ is the thermal conductivity of the particle, and $\mathbf{n}$ is the normal vector at the particle surface where $|\mathbf{x}| = a$.
The heat transfer coefficient $h$ is given by
\begin{equation}
   h = h_0\,\alpha\,\frac{p}{T_\text{g}}\,\overline{v_\text{g}} \;  \label{eq:h}
\end{equation}
\citep{Rohatschek1985}, where $p$ is the local pressure, and the constant $h_0$ is $1/2$ for monatomic gases (used here), and $3/4$ for diatomic gases.
The thermal accommodation coefficient $\alpha$, which is usually of order unity, is assumed to be unity below.
$\overline{v_\text{g}}$ denotes the mean of the magnitude of the three-dimensional thermal speed of the gas molecules or atoms, given by
\begin{equation}
   \overline{v_\text{g}} = \sqrt{\frac{8 R_\text{g} T_\text{g}}{\pi M_\text{g}}} \; . \label{eq:v_gdef}
\end{equation}
We assume a molar mass $M_\text{g} = 2.34 \, \mathrm{g/mol}$. 
$R_\text{g}$ is the universal gas constant.
Employing a separation of variables technique, the solution to the heat transfer problem in \eqref{eq:xfer} for a particle can be expressed as a series 
\begin{equation}
	T_\text{d}(\mathbf{x},t) = \sum_{n=1}^{\infty} g(\mu_n, \mathbf{x}) \, \mathrm{e}^{-t/\tau_n}\;. \label{eq:Tdtseries} 
\end{equation}
The time constants $\tau_n$ for a spherical particle of radius $a$ are given by
\begin{equation}
	\tau_n = \frac{c_p\, \rho_\text{d}}{k}\frac{a^2}{\mu_n^2} \; . \label{eq:taun}
\end{equation}
The values $\mu_n$ are the strictly monotonically increasing roots of the implicit transcendental equation
\begin{equation}
	\frac{h~a}{k} = 1-\mu_n \cot\mu_n \;. \label{eq:condition}
\end{equation}
The function $g$ describes the dependence on the spatial coordinates $\mathbf{x}$ and will not be considered further.
For $t>t_0=0.2 \, c_p\, \rho_\text{d} \, a^2/k$ the $n=1$ term of \eqref{eq:Tdtseries} dominates the heating \citep{Incropera2002_fundamentalsHeatTransfer}, so the temperature evolution of the dust particle can be approximated by
\begin{equation}
	T_\text{d}\simeq \left(T_\text{w}-T_\text{c}\right)\left(1-\mathrm{e}^{-t/\tau_1}\right)+T_\text{c} \; . \label{eq:Td}
\end{equation}  

In order to compute the time constant $\tau_1$, we must find the root $\mu_1$ of \eqref{eq:condition}.
To do this, we use a root finder for varying values of $p, r, T_g,$ and $k$ within the ranges
  $p=10^{-4}$--$10^2\,\mathrm{Pa}$, $r=10^{-3}$--$1\,\mathrm{mm}$,
  $T_\text{g}=50$--$2000\,\mathrm{K}$ and
  $k=10^{-3}$--$7.5\,\mathrm{W\ m^{-1} \ K^{-1}}$.
The value $\tau_1$ can be approximated to within a factor of two for 240,000 combinations of parameters with the empirical equation
\begin{equation} \label{eq:tau1}
	\tau_1=\frac{0.214\, a^2 \, h+k \, a}{3 h} \, \frac{c_p\, \rho_\text{d}}{k}\;.
\end{equation}
\eqref{eq:tau1} allows us to establish whether $\tau_1$ is short.
For radii of 10 $\rm \mu m$ and 1 mm at 1 Pa pressure ($c_p=500\,\mathrm{J~kg^{-1}~K^{-1}}$,
$\rho_\text{d}=3000\,\mathrm{kg~m^{-3}}$, $T_\mathrm{g}=1500\,\mathrm{K}$) \eqref{eq:tau1} yields $\tau_1 = 4\,\mathrm{s}$ and $\tau_1 = 400\,\mathrm{s}$, respectively.
Therefore, for sub-mm-sized particles, gas and particles are thermally well coupled on short timescales.

As the particles heat up, the background radiation field temperature rises.
For simplicity we assume the whole ensemble heats at the same rate.
Extreme cases of large particles at very low-pressure (equivalent to large vertical height) are unlikely due to sedimentation. 
In any case, even if we consider only millimeter-sized objects at moderate pressures, the solids heat up on timescales of a minute.
We therefore consider the temperature fluctuation in the gas to be a temperature fluctuation in the solids as well.
Hence, our proposed drift mechanism driven by temperature fluctuations still works if the gas is heated as the gas quickly transfers this heat to the solids.

\section{Radiation Pressure and Photophoresis}\label{sec:rpp}

As a first toy model to describe a temperature fluctuation that can exert radiation pressure and  photophoretic forces, we take two infinite plane-parallel black-bodies with temperatures $T_\text{c}$ and $T_\text{w}$.
The radiative flux from each side follows the Stefan-Boltzmann law. The net radiative flux density (hereafter simply referred to as flux) on a particle located in between the layers is
\begin{equation}
	I = \epsilon \, \sigma_\text{SB} \, \left({T_\text{w}}^4 - {T_\text{c}}^4\right) \; , \label{eq:intensity_temperature_step}
\end{equation}
where $\sigma_\text{SB}=5.67\times 10^{-8}\,\mathrm{W\,m^{-2}\,K^{-4}}$ is the Stefan-Boltzmann constant and $\epsilon$ the emissivity, assumed unity.
\begin{figure}
	\includegraphics[width=0.5\textwidth]{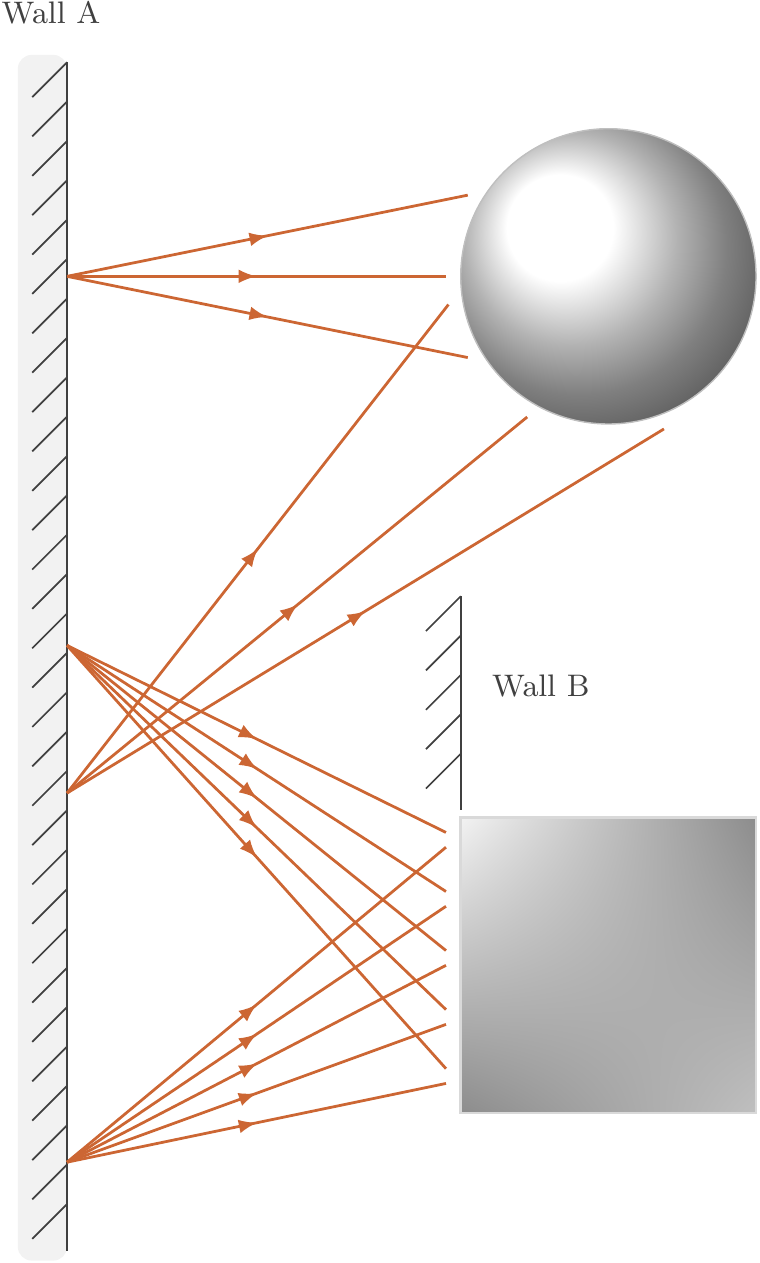}
	\caption{The orange lines represent the radiative flux $I\propto\sigma_\text{SB}T_\text{d}^4$.}
	\label{fig:sketch}
\end{figure}

We note that the following considerations have in mind to approximate the photophoretic force. The literature on photophoresis regularly treats radiation impinging on a particle as being perfectly one directionional and calls it intensity.  This might not fully agree to the astrophysical use of the term intensity. From units the important quantity rather corresponds to a flux  $\rm (W/m^2)$. 

This is only the first paper to outline the idea. Any full 3d treatment is beyond its scope. Therefore, we intend to give only an effective 1-dimensional directional radiative ``flux'' to calculate the photophoretic force. To do so, we simplify the treatment in regions of different temperatures in a model as sketched in the following. 
If we consider a cube particle instead of a sphere, then per definition the side of the cube facing a warm wall (A) (\figref{fig:sketch}) will receive a flux of $I\propto\sigma_\text{SB}T_\text{d}^4$. We might just place the cube right onto a hot wall (B) to make this clear. The touching side gets exactly that flux.
On the other hand, every side of the cube will still get radiation from the wall as a sphere does but due to symmetry every part of the sides gets the same radiation as each area sees a half space of the infinite wall. While this radiation would act as a ``flux'' onto the cube and not average to zero it would not result in a photophoretic force as the sides are heated equally. This does not contribute to the force pointing away from the wall. Only the area in contact with the wall is important for photophoresis. Radiation onto the sides increases the average temperature but does not produce a photophoretic force by itself.
This consideration qualifies approximations for the photophoretic force for directed illumination to be applied in this specific problem. Therefore, we use the term flux in this paper marked as $I$ ($F$ is already taken for force). As the forces from both sides add up the important quantity to calculate photophoretic forces is the difference between the two ``fluxes'' from two opposing directions as e.g. given above for the case of two optical thick walls of different temperatures.


Particles subject to such non-isotropic radiation experience radiation pressure. 
Assuming that the particle absorbs all incident radiation $I$ and approximating the radiation to be perpendicular to the planes, the force exerted on the particle by the radiation 
\begin{equation}
	F_{\text{rp}} = \frac{\pi}{c}\,  a^2\,  I \;, \label{eq:Frp}
\end{equation}
where $a$ is the particle radius and $c$ the speed of light.
The particle accelerates until the force is balanced by the gas drag at a velocity
\begin{equation}
	v = \frac{F}{m} \, \tau_{v} \;, \label{eq:driftvel}
\end{equation}
where $\tau_{v}$ is the gas-grain coupling time and $m$ is the particle mass. In the free molecular flow regime, applicable here, where the mean free path of the gas molecules exceeds the particle size \citep{Blum1996},
\begin{equation}
	\tau_{v} = 0.68 \, \frac{m}{A} \, \frac{1}{\rho_\text{g} \, \overline{v_\text{g}}} \label{eq:tauv}
\end{equation}
where $A=\pi a^2$ is the particle cross section, $\rho_\text{g}$ is the gas density, and $\overline{v_\text{g}}$ is the mean gas thermal velocity.
This results in a drift velocity due to radiation pressure of
\begin{equation}
	v_\text{rp} =  \frac{0.68 \, \sigma_\text{SB} \, \left({T_\text{w}}^4 - {T_\text{c}}^4\right)}{c\, \rho_\text{g} \, \overline{v_\text{g}}} \;. \label{eq:vrp}
\end{equation}
This drift velocity from radiation pressure is independent of the particle size or mass and only depends on properties of the disk.
To calculate a typical drift value we assume $T_\text{w} = 1500$ K, $T_\text{c} = 500$ K, $\rho_\text{g} = 10^{-6}\,\mathrm{kg\ m^{-3}}$, $\overline{v_\text{g}} = 3$ km~s$^{-1}$ (appropriate for an intermediate gas temperature $T_g = 1000\,$K), to find $v_\text{rp} = 21\,\mathrm{cm\,s}^{-1}$.
For lower gas densities, radiation pressure velocities can become large, as gas density enters inversely.
While radiation pressure can change the local particle number densities, $v_\text{rp}$ does not  depend on particle size so no size sorting will result.

Photophoretic forces also act on differentially irradiated particles in a gas \citep{paper2,Rohatschek1995,Beresnev1993}.
The photophoretic force for bare and dust-mantled chondrules has been calculated by \cite{loesche2012}, \cite{loesche2013} and \cite{loesche2014} with the latter two works focused on chondrule motion, covering the case of the chondrule temperature being different from the gas temperature.
Here though, after a short equilibration period, the particles and the  gas will have nearly the same temperature ($T_\text{d}\simeq T_\text{g}$).
In this case, a good general expression for the photophoretic force on a homogeneous sphere is \citep{Beresnev1993,paper1}
\begin{align}
	F_\text{ph} &= \frac{\pi}{6} \, \alpha \, \frac{p}{T_\text{g}} \, \frac{a^2 \, I}{k/a+h+4\sigma_\text{SB}\,T_\mathrm{d}^{3}} \label{eq:photopho}
\end{align}
using the same definitions as in \eqref{eq:xfer}.   
\eqref{eq:photopho} is valid in the free molecular flow regime. This is valid for chondrule size particles ($a \sim 1$ mm) if not too close to the star. 

Substituting this force into \eqref{eq:driftvel}, and using \eqref{eq:tauv} for the gas-grain coupling time, as well as substituting for the area $A=\pi\,a^2$ and using the ideal gas law 
\begin{equation}
   p=\frac{R_\text{g}}{M_\text{g}}\, \rho_\text{g}\, T_\text{g} \; , \label{eq:p}
\end{equation}
we get a photophoretic drift velocity
\begin{equation}
	v_\text{ph} =  0.68\frac{R_\text{g}}{M_\text{g}\,\overline{v_\text{g}}} \, \frac{\alpha}{6} \, \frac{ I}{k/a + h + 4\sigma_\text{SB}\, T_\text{d}^3} \;. \label{eq:vph}
\end{equation}
Using the disk and temperature fluctuation values used in our example for radiation pressure ($I$ given by \eqref{eq:intensity_temperature_step}), and typical values for chondrules, $a= 1$ mm, $k=1$ W m$^{-1}$ K$^{-1}$, $\alpha =1$, we find a representative value of $v_\text{ph} = 52\,\mathrm{m\,s}^{-1} \gg v_\text{rp}$.
This drift velocity is also large compared to other gas-grain velocities such as radial inward drift or turbulence induced motion \citep{Weidenschilling1977,Zsom2010}.
\citet{Jacquet2014} suggest drift speeds up to $200\,\mathrm{m\,s}^{-1}$ in shocks, but these velocity 
variations are very short-lived.
The drift velocities calculated here exist as long as a temperature fluctuation is present.

On the one hand, in contrast to radiation pressure drift, the photophoretic drift velocity depends on the particle size $a$.
On the other hand, the drift velocity does not depend on the absolute value of the pressure for large Knudsen numbers (if $h$ is small compared to the radiation term), as both photophoretic force and gas drag increase linearly with gas density. Hence, it is of minor importance if constant pressure or constant density is considered.
The drift velocity also depends on the details of the thermal conductivity of the particle.

For the regime we consider here, the thermal conductivity term $k/a$ dominates the denominator of \eqref{eq:photopho} \citep{Wurm2006}.
Therefore dust aggregates, with thermal conductivities a factor 10 to 100 lower than for chondrules \citep{presley1997}, can reach drift velocities of several hundred meters per second.
These estimates show that photophoretic forces have a significant influence on the motion of particles in a disk with temperature fluctuations.
This can strongly modify the evolution of dust aggregates as bouncing barriers and mass transfer \citep{guettler2010} are no longer well defined, due to the large variations induced by photophoretic force.
\section{Radiation Field}

Since the simple two-temperature toy model of the previous section clearly shows the potential for large effects on particle motion and planetesimal formation, we go one step further here and study the light fluxes, drift velocities, and spatial particle transport at the edge of a \emph{hot spot} temperature fluctuation with a linear temperature gradient.

The basic quantity to compute is the (anisotropic) light flux at a given position in the disk.  We make the approximation that the radiation field is locally in equilibrium with the dust temperature, so that an appropriate starting point is the frequency-averaged grey opacity given by the Rosseland mean opacity, which includes both absorption and scattering effects \citep{1986rpa..book.....R}.
The Rosseland mean opacity $\chi$ of a dust-gas mixture with the canonical interstellar medium dust-gas density ratio $\rho_\mathrm{d}/\rho_\mathrm{g} =$ 0.01 in the temperature range where silicate and carbon grains dominate the opacity is on the order of $\chi = 0.2\,\mathrm{m^2\,kg^{-1}}$ \citep{1994ApJ...427..987B}.
In our disk this translates to an absorption coefficient per unit length $\chi \, \rho_\text{g}$, introducing the optical depth $\tau$ as
\begin{align}
   \mathrm{d}\tau = \chi \, \rho_\text{g} \dd r \; . \label{eq:tau}
\end{align}
However, as the grain size distribution, composition, and dust enhancement over the canonical interstellar medium ratio can reasonably be expected to vary in protoplanetary disks, we take $\chi$ as a parameter in our study in the range of $0.02$--$0.4\,\mathrm{m^2 \ kg^{-1}}$.
We keep the gas density constant at ${\rho_\text{g}}=10^{-6}\,\mathrm{kg\ m^{-3}}$.

At each position we can quantify the light flux incident on a particle.
We consider a one-dimensional problem here, which means that the light flux includes four terms.
First, we divide the light flux into positive and negative directions ($r$) along the temperature gradient.
We calculate the light flux in each direction by considering absorption of radiation and thermal emission from the dust
\begin{equation}
	S(\tau) = \sigma_\text{SB} \, T_\text{d}^4(\tau) \; . \label{eq:source}
\end{equation}
The latter requires that the gas temperature is efficiently transferred to the dust particles, as is
the case for disk locations that are not too optically thin (see estimates above).

Written in terms of the optical depth $\tau$, both light rays obey the following differential equations for the fluxes
\begin{subequations}
	\begin{align}
	0 &= I'_+(\tau) + \left[ I_+(\tau)- S(\tau) \right] \\
	0 &= I'_-(\tau) - \left[ I_-(\tau)- S(\tau) \right]
	\end{align}
\end{subequations}
with their respective solutions
\begin{subequations}
	\begin{align}
	I_{+}(\tau) &= I_{+}(\tau_{0+}) \, \exp{\tau_{0+}-\tau} + \int\limits_{\tau_{0+}}^{\tau} S(\tilde{\tau}) \, \exp{\tilde{\tau}-\tau} \dd \tilde{\tau} \label{eq:Strahl_von_links} \\
	I_{-}(\tau) &= I_{-}(\tau_{0-}) \, \exp{\tau-\tau_{0-}} + \int\limits_\tau^{\tau_{0-}} S(\tilde{\tau}) \, \exp{\tau-\tilde{\tau}} \dd \tilde{\tau} \; . \label{eq:Strahl_von_rechts}
	\end{align}
\end{subequations}
The total flux reads
\begin{align}
I(\tau) &= I_{+}(\tau) - I_{-}(\tau) \; . \label{eq:intensity}
\end{align}
The source function $S$, the radiative flux $I$, the photophoretic force $F_\text{ph}$ and the drift speed $v_\text{ph}$ are linear in the average $\epsilon$, and for simplicity we assume $\epsilon=1$ further on.

\section{Hot spot model}

We consider a temperature profile that decreases linearly over a spatial distance $l$ from a temperature of $T_\text{w} = 1500$ K to a temperature of $T_\text{c} = 500$ K (we keep both values fixed in this paper)
\begin{equation}
   T_\text{d/g}(r,l) = \begin{cases}
      T_\text{w} & r \le 0 \\
      T_\text{w}-\left(T_\text{w}-T_\text{c}\right)\,r/l & 0\le r\le l \\
      T_\text{c} & r \ge l
   \end{cases} \; . \label{eq:temp}
\end{equation}
We do not consider the time evolution of these fluctuations yet but keep the profile constant.
The spatial distance $l$, the first characteristic scale of this problem, is varied and the results compared to the most extreme case ($l=0$, a step function), used in our toy model.
As mentioned earlier, for a simplification of the following discussion, we restrict ourselves to spatially constant values $\rho_\text{g}(r)=\mathrm{const}$ and $\chi(r)=\mathrm{const}$.

Figure~\ref{fig:results} shows an example of such a temperature profile and the resulting net radiative flux assuming an opacity of $\chi=0.02\,\mathrm{m^2\,kg^{-1}}$, graphically represented by the dotted lines of unit optical depth defined through the second characteristic scale $\Delta$, defined by (compare \eqref{eq:tau})
\begin{align}
	\tau &= \chi\,\rho_\text{g}\,r\;, \label{eq:tau_chi} \\
	\Delta &\equiv \left(\chi\,\rho_\text{g}\right)^{-1} = |r|/\tau \; . \label{eq:Delta}
\end{align}
As can be seen, the edge of the radiative flux is smoothed, compared to the sharp transition of the temperature fluctuation.

Written in terms of optical depth $\tau$ instead of $r$, the temperature (\eqref{eq:temp}) has the form $T_\text{d/g}=T_\text{d/g}(\tau,l/\Delta)$ (following \eqref{eq:Delta}).
The principal idea is, that obviously different configurations of the two characteristic scales $l$ and $\Delta$ with the same ratio yield the same linear temperature ramp $T_\text{d/g}(\tau,l/\Delta)$ in $\tau$.
In our system the source function $S$ (\eqref{eq:source}) can therefore be written as a function of $\tau$ and $l/\Delta$, too; as can the flux $I$ (\eqref{eq:intensity}).
Indeed, as we are assuming a spatially constant gas density $\rho_\text{g}$, the pressure $p$ (\eqref{eq:p}), the thermal speed $\overline{v_\text{g}}$ (\eqref{eq:v_gdef}), and hence the heat transfer coefficient $h$ (\eqref{eq:h}) are also only functions of $\tau$ and $l/\Delta$.
It follows that $F_\text{ph}$ (\eqref{eq:photopho}) and $v_\text{ph}$ (\eqref{eq:vph}) can be written as functions only of $\tau$ and $l/\Delta$, and of parameters such as $a$, $\alpha$ and $k$.

\figref{fig:flux2} shows the (spatial) maximum light flux
	\begin{equation}
		I_{\max}(l/\Delta)=\smash{\displaystyle\max_{\tau}}\,I(\tau,l/\Delta) \; , \label{eq:Imax}
	\end{equation}
	across the temperature drop over $l/\Delta$.
	The maximum light flux in the case of a temperature step $(l = 0)$ is marked as a dotted line.

\begin{figure}
	\includegraphics[width=1\columnwidth]{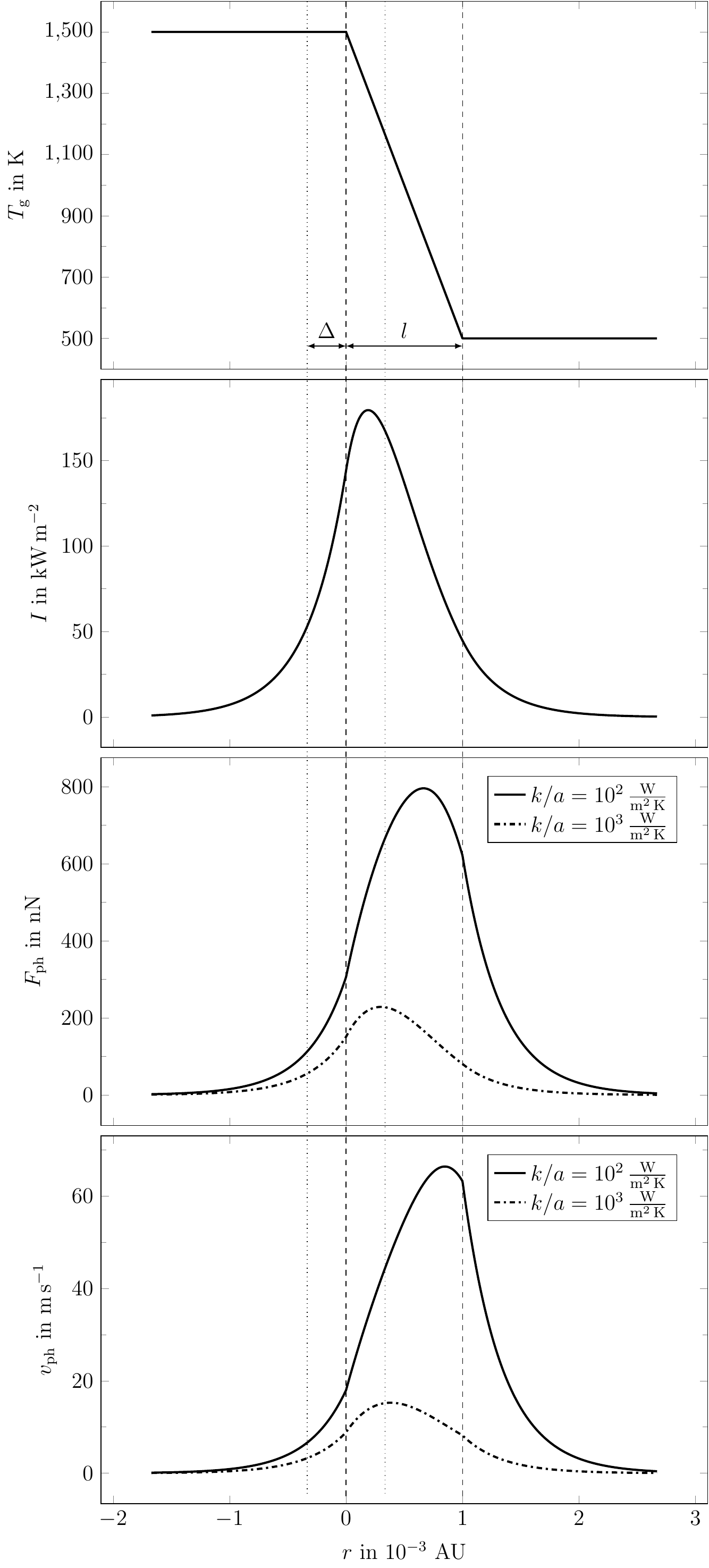}
	\caption{\label{fig:results}
		First: linear temperature ramp of $l=10^{-3}\,\mathrm{AU}$ to model the temperature fluctuation.
		Second: resulting radiative flux for this ramp at $\chi=0.02\,\mathrm{m^2\,kg^{-1}}$, $\rho_\text{g}=10^{-6}\,\mathrm{kg\,m^{-3}}$.
		Dashed and dotted lines show the two characteristic scales $l$ and $\Delta$, respectively.
		Third and fourth: photophoretic force (\eqref{eq:photopho}) and photophoretic drift speed  (\eqref{eq:vph}) for a dust-like (lower $k/a$) and a chondrule-like (higher $k/a$) particle each.
		The maximum values of radiative flux $I$, force and drift speed do not necessarily coincide.
		}
\end{figure}
These flux variations can be translated to photophoretic forces.
In the next section, the calculated forces are used to quantify the drift velocities for the case of $l > 0$.

\begin{figure}
   \includegraphics[width=1\columnwidth]{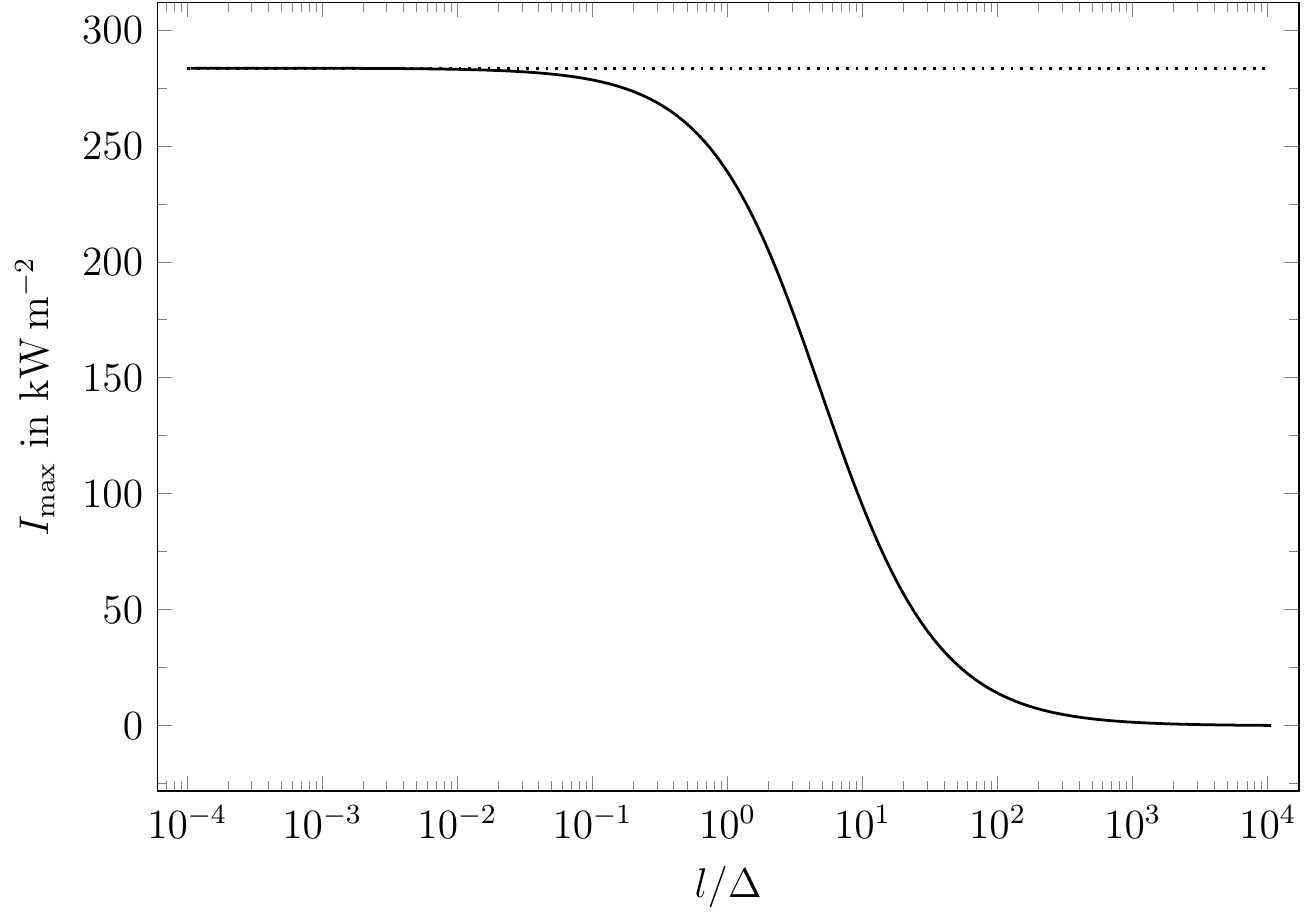}
   \caption{\label{fig:flux2}Maximum light flux depending on $l/\Delta$. The dotted line marks the maximum light flux for a temperature step, determined by \eqref{eq:intensity_temperature_step}.}
\end{figure}
\section{Drift}

The photophoretic forces computed in the last section lead to particle motion away from the hot spot with a velocity given by \eqref{eq:vph}.  
After some time, particles initially situated along the edge of the fluctuation will be trapped outside the hot spot.
In this case, the transition zone will be cleared of particles and the particle concentration will be enhanced outside of the hot spot.

Photophoretic drift is size dependent.
Therefore, with time, large particles will move further away from the inner edge of the hot spot than smaller ones.
To quantify this effect, we used \eqref{eq:vph} to compute the maximum velocities of four combinations of particle parameters here: chondrule-like particles with a radius of 1 mm or of 0.1 mm and thermal conductivity of $k = 1 \,\mathrm{W\ m^{-1} \ K^{-1}}$, and dust aggregates of the same radii, but with a thermal conductivity of $ 0.1 \,\mathrm{W\ m^{-1} \ K^{-1}}$.
Their computed (spatial) maximum drift speeds, i.e.
\begin{equation}
	v_{\text{ph,}\max}(l/\Delta)=\smash{\displaystyle\max_{\tau}}\,v_{\text{ph}}(\tau,l/\Delta) \; ,  \label{eq:vmax}
\end{equation}
are shown in \figref{fig:vph} and \tabref{tab:table1} across the temperature drop over $l/\Delta$.

\begin{figure}
	\includegraphics[width=1\columnwidth]{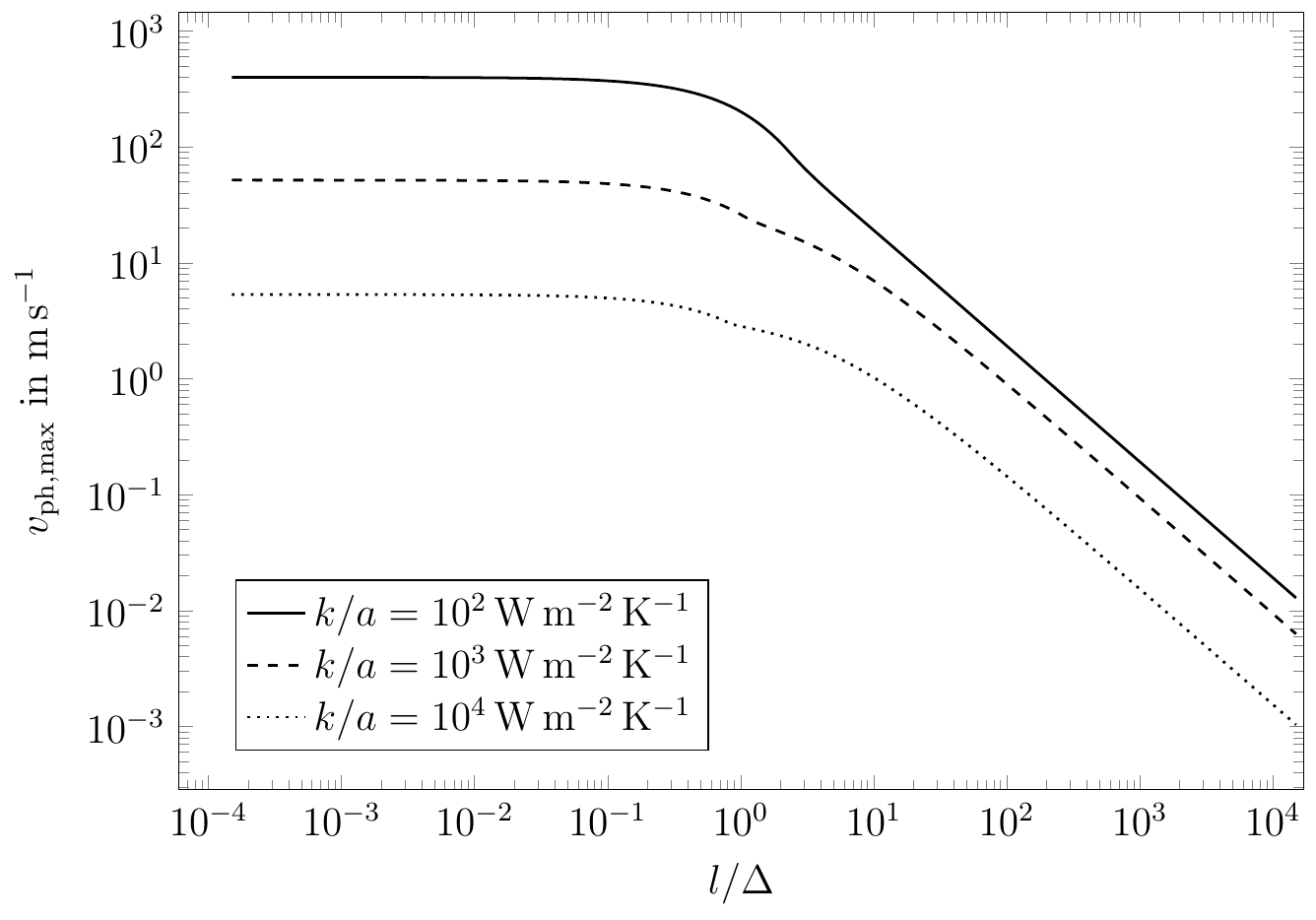}
	\caption{
		\label{fig:vph}Maximum drift speed using \eqref{eq:vph}. The place of the maximum speed is not necessarily coincident with the place of maximum radiative flux (\eqref{eq:photopho}), due to the varying gas temperature. To determine $h$, a density of ${\rho_\text{g}}=10^{-6}\,\mathrm{kg\, m^{-3}}$ was used. The maximum values for all three different particles are 400, 52 and $5\,\mathrm{m\, s^{-1}}$, respectively. For $l/\Delta=1$, drift speeds of 203, 26 and $2.8\,\mathrm{m\, s^{-1}}$ are achieved. Also see \tabref{tab:table1}.
		}
\end{figure}

\begin{table}
	\centering
	\caption{\label{tab:table1}Maximum drift velocities depending on particle properties and temperature fluctuation for fixed $T_\text{w}$ and $T_\text{c}$.}
	\begin{tabular}{L{1.5cm} C{1.2cm} C{1.2cm} C{1.6cm} >{\bfseries}C{1.2cm}} \toprule
		\multicolumn{1}{c}{particle} & \multicolumn{2}{c}{configuration} & \multicolumn{2}{c}{$v_\text{ph,max}(l/\Delta)$ in $\mathrm{m\, s^{-1}}$} \\
		\cmidrule (l){2-3} \cmidrule (l){4-5}
		 & $k$ & $a$ & \multicolumn{2}{c}{$l/\Delta$} \\
		 & $\rm W\,m^{-1}\,K^{-1}$ & $\mathrm{mm}$ & $10^{-4}$ & $\boldsymbol{1}$ \\
		\midrule
		Chondrule	& 1.0	& 1.0	& 52 	& 26 \\
		Chondrule	& 1.0	& 0.1	& 5 	& 2.8 \\
		Dust		& 0.1	& 1.0	& 400 	& 203 \\
		Dust		& 0.1	& 0.1	& 52 	& 26 \\
		\bottomrule
	\end{tabular}
\end{table}
 
Large dust aggregates move the fastest. At millimeter size, they can reach velocities exceeding several 100~m~s$^{-1}$.
Small chondrules move the slowest, only reaching 5~m~s$^{-1}$.
As implied by \figref{fig:results}, significant fractions of these maximum speeds can be reached across the entire transition region.
Considering chondrules of 1 mm and 0.1 mm the relative velocity at maximum, close to the warm side of the edge are several tens of meters per second (\tabref{tab:table1}).
Therefore, locally, particles will be separated by size.
After only an hour, small and large particles are already separated by 100 km in this simple sketch, which might be enough to place them into different parent bodies or different parts of a parent body -- size sorted -- later on.
This also implies that recently formed chondrules can escape a region of a few thousand kilometers within a day and might cool at a rate characteristic of observed chondrules \citep{Connolly2006}.

The estimates of drift velocities for dust aggregates exceeding 100 m~s$^{-1}$ are also intriguing. For collisional evolution, such relative velocities depending on size are important. 
These velocity magnitudes for millimeter-sized particles will be dominant over other velocities, both absolute and relative. 
So even for smaller temperature fluctuations, photophoretic drift could have a significant impact, changing the particle sizes at which particles are expected to reach the bouncing and fragmentation barriers.

Obviously, the drift speed $v_\text{ph}$ for dust is much higher than for chondrules.
Hence, in an environment, where dust and chondrules coexist, dust clumps will collide with chondrules much more often than chondrules with each other, which could promote formation of chondrules with extreme dust rims \citep{1992GeCoA..56.2873M} and igneous rims that are melted dust layers \citep{2008M&PS...43.1725E}.

This could alter the picture of the first phases of collisional evolution quite a bit.
Bouncing barriers have been discussed in recent years \citep{guettler2010,Zsom2010,Kelling2014}. 
These occur when particles reach a size and collisional velocity regime where growth of aggregates is stalled.
This occurs for millimeter-sized silicate particles.
At this size, bouncing collisions dominate and inhibit further growth.
However, bouncing collisions are mostly restricted to velocities below 1~m~s$^{-1}$.
In a disk with temperature fluctuations, collision velocities can reach much larger values locally and the bouncing barrier can locally be overcome. 

Even more, though, the distribution of velocities at different sizes might render the very notion of a bouncing barrier questionable.
\citet{drazkowska2014} and \citet{windmark2012} show that the bouncing barrier can be overcome if a few larger seeds are embedded.
If the collision velocities exceed 1~m~s$^{-1}$, mass is transferred from small to large particles instead of bouncing.
Photophoretic drift at the edge of a temperature fluctuation will locally raise the velocity   dispersion sufficiently that mass transfer rather than bouncing becomes dominant, and thus could provide larger seeds for further growth in the quiet disk after a large temperature fluctuation vanished. However, as the peak velocities are rather large, growth is not guaranteed \citep{Meisner2013}.
Therefore, we leave the speculation at this point: although the effects might be paradigm changing, the details can only be elucidated by future modeling and experiment. 

\section{Caveats}
An aspect of concern raised in earlier works on photophoresis is particle rotation.
If particles are initially in a random state of rotation this has influence on the photophoretic strength, as rotation impedes the establishment of the strongest possible temperature gradients across the particle.
This has been studied in \cite{loesche2013,loesche2014}.
However, random rotations damp on timescales of the gas grain friction time, while photophoresis itself only excites rotations around the direction of radiation, which do not exchange the front and back side \citep{vanEymeren2012}.
Therefore, particle rotation does not impede photophoretic motion. 

In this work, we have applied strictly one-dimensional approximations of the radiative transfer problem, and considered the net flux of the radiation field to be the one-sided illumination that appears in the description of the photophoretic force on a particle.
This approximation neglects the full nonlinearity of the differential illumination of the front and 
rear of the particle, caused by the rays striking the side and rear of the particle from forward angles, and vice-versa.
The effect of our one-dimensional approximation should be uniformly to yield a larger implied temperature contrast between the front and rear of the particle, as compared to a genuinely three-dimensional approach. 
In turn, this yields a higher photophoretic force in the one-dimensional treatment used in this work than a fully three-dimensional radiation field.
A subsequent work in this series will treat the fully three-dimensional problem in an optically thick medium.
\section{Conclusion}

Starting with \citet{Krauss2005} and \citet{Wurm2006} the photophoretic motion of particles in protoplanetary disks has been discussed in a number of papers.
While the first works considered the idea in principle, and assumed an overall optically thin disk, this was because the only heating mechanism considered was starlight, which of course can not penetrate optically  thick disks.
Later work then concentrated on the inner edge or surface of the disk \citep{wurm2009,Wurm2010, Kelling2013} or transition disks \citep{Krauss2007,Mousis2007,Herrmann2007,Cuello2016}. 

However, we show here that an optically thick disk can have photophoretic forces acting within its interior.
We have discussed the example of a disk with temperature fluctuations, as suggested by the requirements for chondrule formation.
We find that photophoretic motions in this case can be very fast, reaching velocities $>100$~m~s$^{-1}$, and dominating the zoo of different drift mechanisms for dust aggregates. 

If temperature fluctuations lead to chondrule formation, the resulting particles can rapidly be pushed to cooler regions afterward.
They can locally be size sorted and concentrated, which might lead to planetesimal formation
shortly after their formation, e.g. by gravitational instabilities \citep{johansen2007,cuzzi2008,chiang2010,dittrich2013}.

In any case the anisotropy of thermal radiation induces a hitherto undiscussed relative velocity component between particles. This might change the early phases of growth of dust aggregates.
Evolving dust aggregates can experience large velocity differences   that can dominate the collision kernel in growth scenarios. 
Details are subject to details of photophoresis acting on different particles.
A first example study of these details is work by \citet{Kuepper2014b}, who recently started to quantify the photophoretic properties of aggregates.  

We focused here on the high-temperature case required to explain chondrule formation.
However, photophoretic motions remain significant at lower temperatures.
In a certain sense this scheme was already applied to more specific cases of systematic temperature differences by \cite{teiser2013}.
They discuss photophoretic particle motion in the context of giant planet formation where a local heat source is present.
The application to a fluctuating disk goes far beyond that, however.

To summarize, photophoretic forces might be far more than a subtle detail in specialized situations at the edge of protoplanetary disks.
Instead these non-equilibrium effects in low pressure gaseous environments might govern aspects of
particle transport and evolution even in the most optically thick parts of protoplanetary disks.

\section*{Acknowledgements}

C.L. was funded by DFG 1385 and a Kade Fellowship (AMNH).
We thank Colin P. McNally, Mordecai-Mark Mac Low and Alexander Hubbard for many constructive discussions.




\bibliographystyle{mnras}
\bibliography{references} 




\appendix
\section{Notation}
\begin{table}
	\centering
	\caption{\label{tab:table2}Notation.}
	\begin{tabular}{ >{$}r<{ $} L{\columnwidth}  }\toprule		
		\text{variable} & meaning 	\\
		\midrule
		
		a				& radius of spherical particle (dust, chondrule) \\
		c_p				& isobaric heat capacity of dust	\\
		\chi			& Rosseland mean opacity \\
		\Delta			& width, where $\tau=1$ (in the same units as $r$), \eqref{eq:Delta} \\
		\epsilon		& (mean) emissivity, considered unity \\
		F_\text{rp}		& radiation pressure force, \eqref{eq:Frp} \\
		F_\text{ph}		& photophoretic force, \eqref{eq:photopho} \\
		g=g(\mu_n,\mathbf{x})	& spatial part in the expansion of the dust temperature $T_\text{d}$, \eqref{eq:Tdtseries} \\
		h				& heat transfer coefficient, \eqref{eq:h} \\
		I				& radiative flux areal density, \eqref{eq:intensity} \\
		I_{\max}		& maximum of $I$ in $\tau$ or $r$, \eqref{eq:Imax} \\
		I_+				& ray towards $r$, \eqref{eq:Strahl_von_links} \\
		I_-				& ray towards $-r$, \eqref{eq:Strahl_von_rechts} \\
		k				& thermal conductivity of (dust) particles	\\
		l				& temperature step width (in the same units as $r$) \\
		M_\text{g}		& molar mass of gas \\
		\mu_n			& roots of \eqref{eq:condition} \\
		\mathbf{n}		& normal vector of a surface \\
		p				& gas pressure, we consider the ideal gas law, \eqref{eq:p} \\
		r				& radial distance in disk \\
		R_\text{g}		& universal gas constant \\
		\rho_\text{d}	& dust mass density	\\
		\rho_\text{g}	& gas mass density	\\
		S				& source function, \eqref{eq:source} \\
		\sigma_\text{SB}& Stefan-Boltzmann constant \\
		t				& time \\
		T_\text{d}		& dust temperature	\\
		T_\text{g}		& gas temperature	\\
		T_\text{c}		& lower gas temperature (linear temperature ramp)	\\
		T_\text{w}		& higher gas temperature (linear temperature ramp)	\\
		\tau			& optical depth, we consider $\tau\in\left(\tau_{0-},\tau_{0+}\right)$, \eqref{eq:tau} \\
		\tau_n			& time constant for dust particle heat-up in $n$-th order, \eqref{eq:taun} \\
		\tau_v			& gas-grain coupling time, \eqref{eq:tauv} \\
		\overline{v_\text{g}} & $=\overline{|\mathbf{v}_\text{g}|}$, i.e. mean of the thermal speed of the gas, \eqref{eq:v_gdef} \\
		v				& drift speed of suspended particle, balanced by gas drag, \eqref{eq:driftvel} \\
		v_\text{rp}		& drift speed caused by radiation pressure, \eqref{eq:vrp} \\
		v_\text{ph}		& drift speed caused by photophoresis, \eqref{eq:vph} \\
		v_\text{ph,max}	& maximum of $v_\text{ph}$ in $\tau$ or $r$, \eqref{eq:vmax} \\
		\mathbf{x}		& spatial coordinate (used for the heat transfer problem)	\\
		
		\bottomrule
	\end{tabular}
\end{table}


\bsp	
\label{lastpage}
\end{document}